# Transformational application of Artificial Intelligence and Machine learning in Financial Technologies and Financial services: A bibliometric review

Vijaya Kanaparthi
Microsoft
Northlake, Texas, USA # 76226
datalivesite@gmail.com

*Abstract*—In this study, I employ a multifaceted comprehensive scientometric approach to explore the intellectual underpinnings of AI and ML in financial research by examining the publication patterns of articles, journals, authors, institutions, and nations by leveraging quantitative techniques, that transcend conventional systematic literature reviews, enabling the effective analysis of vast scientometric and bibliographic data. By applying these approaches, I identify influential works, seminal contributions, thought leaders, topical clusters, research streams, and new research frontiers, ultimately fostering a deeper understanding of the knowledge structure in AI and ML finance research by considering publication records from 2010 to 2022 from several search engines and database sources. The present study finds a marked increase in publications from 2017 to 2022, which highlights a growing interest and expanding research activity in the field, indicating its potential significance and relevance in the contemporary academic landscape.

## I. Introduction

The swift expansion of artificial intelligence (AI) and machine learning (ML) technologies across various industries has prompted an increased interest in the application of these techniques within the financial sector [1]. Consequently, the integration of AI and ML has produced a diverse range of implications for financial services, including asset allocation, risk management, and market analysis. The continuous expansion of digital data in finance presents both challenges and opportunities for AI and ML applications [2]. As financial institutions increasingly rely on alternative data sources, such as social media, news articles, and satellite imagery, the need for innovative techniques to manage and analyze this information becomes paramount [3]. AI and ML models can effectively process large, unstructured, and high-dimensional data, enabling the extraction of valuable insights and the discovery of intricate patterns that may elude traditional econometric methods.

There is a need for transparent and accountable decision-making processes in financial institutions since they are held to strict regulatory standards [4]. The adoption of interpretable AI and ML techniques, such as LASSO regression, decision trees, and rule-based systems, can facilitate a better understanding of the underlying mechanisms driving financial predictions and support more informed decision-making. Another promising area of research is the integration of AI and ML with network analysis and complex systems theory to better capture the complex relationships between financial entities and enhance their predictive capabilities [5]. Yet another area of AI and ML research is Natural language processing (NLP) which has significant potential for fi-nance applications [6]. The analysis of textual data, such as financial reports, news articles, and social media content, can provide valuable insights into market sentiment, investor behaviour, and economic trends. Advances in NLP techniques, such as sentiment analysis and topic modelling, have enabled the development of sophisticated AI and ML models that can effectively process and analyze textual information in finance.

Despite the growing body of literature on this subject, there remains a notable dearth of comprehensive reviews that synthesize the current state of AI and ML research in finance. To address this gap, undertaking an exhaustive bibliometric analysis of the existing literature on AI and ML applications in finance, covering the period from 2010 to 2022. Our study employs bibliometric coupling to identify the key themes, trends, and research directions within the domain of AI and ML in finance. In doing so, the aim to provide a comprehensive overview of the field, as well as establish a foundation for future research endeavours.

The rest of the paper is organized as follows. Section II presents the methodology used in this study. Section III confers the results followed by a discussion in section IV. Section V concludes the paper with some highlights on future scope.

## II. Methodology

### A. Methods

In this study, it employs a multifaceted scientometric approach that leverages quantitative techniques to scrutinize the scientometric and bibliographic landscape [7]. In contrast to conventional systematic literature reviews, a scientometric examination is well-equipped to address domains permeated with vast quantities of scientometric and bibliographic data. Specifically, adhering to the four-pronged strategy for conducting scientometric reviews: (1) delineating the objectives and scope of the review; (2) selecting the appropriate analytical methods; (3) assembling the data required for analysis; and (4) executing the analysis and presenting the outcomes.

To commence, I establish a clear set of criteria for the inclusion and exclusion of relevant literature of document type "article" and source type "journal". This entails the identifica-tion of pertinent keywords ("data science" OR "deep learning" OR "DL" OR "big data*" OR "automation" OR "natural language processing" OR "NLP" OR "decision tree" OR "support vector machine" OR "SVM" OR "random forest" OR "RF" OR "gradient boosted tree" OR "Naive Bayes" OR "artificial neural network" OR "ANN" OR "ML classification" OR "learning model" OR "Nearest Neigh-bour" OR "big data analytic" OR "reinforcement learning" OR "text classification" OR "Bayesian" OR "machine learning pipeline" OR "DNN") AND ("finance" OR "financial service" OR "fintech" OR "financial technology"), subject categories ("business, management, and accounting", "economics, econometrics, and finance"), screening language (English), and time frames (2010-2022), which will enable a compre-hensive yet focused review. Additionally, utilizing multiple databases and search engines, such as Web of Science, Scopus, and Google Scholar, to ensure that the data collect-ed is both exhaustive and representative of the field's intellectual breadth.

*B. Data collection*

In line to examine the impact of artificial intelligence (AI) and machine learning (ML) on financial management, selecting Scopus as this primary database for data collection. Scopus is deemed the most comprehensive source for peer-reviewed research in finance compared to alternative databases such as Web of Science. Drawing inspiration from a recent literature review [7], employing an extensive set of search terms to ensure a thorough investigation of the subject matter.

I utilize a systematic approach in obtaining a final corpus of 364 articles using VOSviewer, a software tool designed for constructing and visualizing bibliometric networks, providing an insightful exploration of the relationships among documents, au-thors, journals, institutions, and countries for precise topical analysis.

III. RESULTS

Delving into the extensive analysis of AI and ML applications within finance, encompassing critical parameters: citation statistics, authorship, institutional contributions, national trends, and influential papers within the timeframe of 2010-2022. Figure 1 encapsulates this comprehensive analysis. Fig 1(a) visually represents the annual publication count from 2010 to 2022. Commencing at a modest 2 papers in 2010, the trajectory showcases fluctuations but ultimately ascends to 107 publications in 2022, notably surging from 2017 onwards. Fig 1(b) displays a dataset showcasing yearly citation counts. The data delineates a non-monotonic pattern, showcasing peaks like the substantial 569 citations in 2017 and a zenith of 958 citations in 2019. However, a decline to 290 citations in 2022 hints at a potential downward trend. Fig 1(c) elucidates the publication records of prominent authors. Wang Y. and Zhang H. lead with five papers each, garnering 91 and 63 citations, respectively. Other significant contributors include Li H., Sun J., and Li Z., reflecting both substantial paper outputs and received citations. Fig 1(d) scrutinizes the research contributions of institutions. The University of Oxford leads with six publications, closely followed by Renmin University of China with five. Noteworthy institutions exhibit parity with four papers each. Fig 1(e) delineates the distribution of research output across countries. China emerges as the leader with 76 publications, trailed closely by the United States (72) and the United Kingdom (55). Other contributors include India (24), Australia (19), France (17), Germany (14), Taiwan (12), Hong Kong, and Spain (11 each). Fig 1(f) presents citation numbers for the top 20 research articles. The most cited paper [8] holds the highest citation count. These findings encapsulate the dynamic evolution of AI and ML in finance, marked by burgeoning publications and citations, diverse contributions from global authors and institutions, and varying research productivity among nations.

Moving into the realm of finance research, distinct machine learning algorithms stand out due to their prevalence and application. Random Forests (RF) emerge prominently across diverse domains, tackling topics ranging from financial performance to risk management and credit assessment. Support Vector Machines (SVM) play a pivotal role in risk assessment and credit research, while Neural Networks find applications in bankruptcy prediction, credit risk assessment, and systemic risk analysis. Though less prevalent, deep learning techniques have been employed in behavioural finance, insurance, and pricing models. Conversely, specific machine learning algorithms witness limited utilization within finance research. Algorithms such as Naive Bayes, K-Nearest Neighbors (KNN), and Gradient Boosting feature sparingly and only in select topics. Notably, areas like financial intermediation and portfolio management display minimal or no use of machine learning techniques, suggesting unexplored avenues for future research, as illustrated in Fig. 2. This analysis underscores the widespread adoption of machine learning methodologies such as RF, SVM, and Neural Networks in finance research, simultaneously shedding light on the underutilization of alternative techniques. These findings serve as a valuable reference for researchers aiming to comprehend the contemporary landscape of computational finance. Additionally, they offer insights to explore lesser-utilized machine learning methods within specific finance domains for prospective investigations.

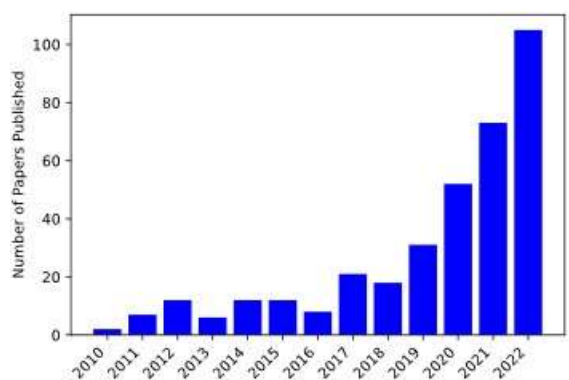

(a)

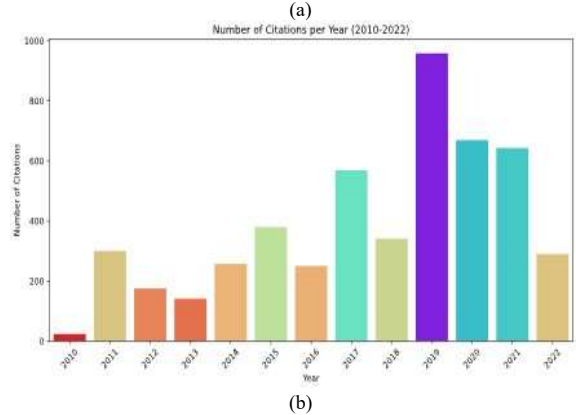

(b)

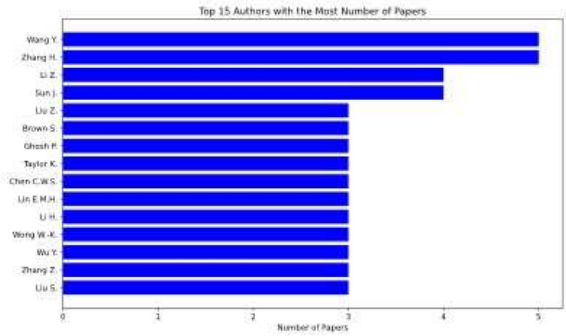

(c)

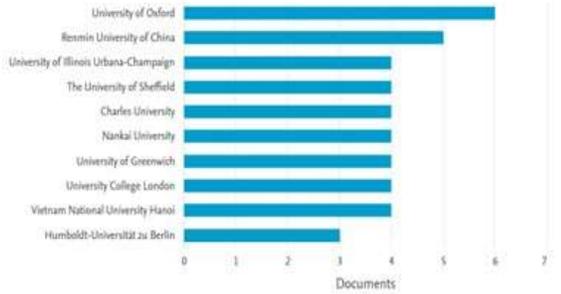

(d)

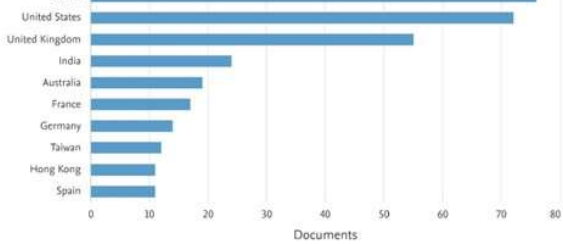

(e)

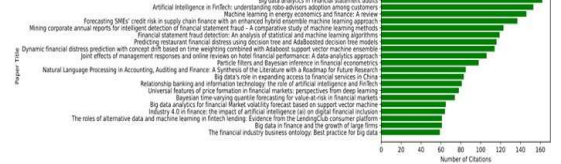

(f)

(a) Yearly publication status from 2010 to 2022 in the field of AI & ML in finance; (b) Citation statistics of the AI & ML papers in the field of finance from 2010 to 2022; (c) Top authors with highest number of papers published during 2010 to 2022 period; (d) Institution-wise publication count for the study period – 2010 – 2022; (e) Country-wise publication numbers during the study period; (f) Most cited papers in the field of AI/ML in finance.

Fig. 1. Result Analysis of Publication Activity of AI & ML Application on Finance

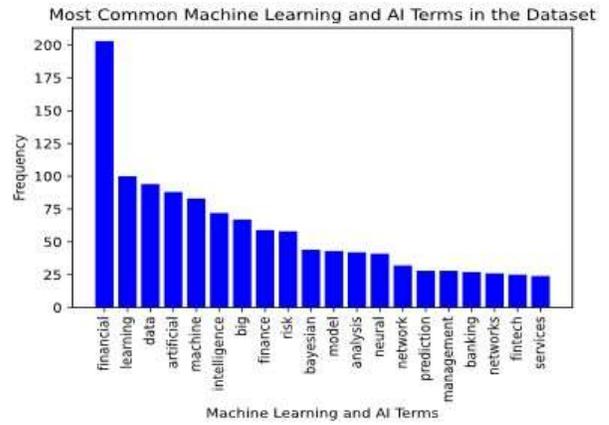

Fig. 2. Most frequently used terms in the field of finance research

## IV. DISCUSSION

### A. Financial fraud detection

Financial fraud detection has advanced significantly, most notably via the use of supervised learning approaches such as support vector machines (SVM) and neural networks, which have shown successful in detecting credit card fraud, insider trading, and money laundering [9], [1]. In addition, unsupervised learning approaches such as clustering and outlier identification have gained popularity for detecting elusive fraud trends inside complex datasets [1], [10]. The rising tendency is to use deep learning models such as recurrent neural networks (RNNs), long short-term memory (LSTM) networks, and convolutional neural networks (CNNs) to represent complicated connections in financial data [11]. Integrating natural language processing (NLP) tools, particularly in evaluating textual material (e.g., bank statements, emails, and social media postings), is a critical step in this arena, providing insights into prospective fraudsters' goals and motives [12]. Researchers investigate feature-level, decision-level, and model-level fusion methodologies in order to construct robust models capable of identifying fraudulent behaviours across a wide range of data kinds and sources [13]. Furthermore, the use of distributed computing frameworks such as Hadoop and Spark has simplified the management of massive data streams, enabling real-time processing and enabling financial institutions to vigilantly monitor and minimise fraud threats.

### B. Stock prediction and sentiment analysis

Recent years have seen great breakthroughs in sentiment analysis, owing to the combination of machine learning and natural language processing approaches. Researchers and practitioners are increasingly using these approaches to anticipate stock movements and analyse sentiment [14]. The emergence of different methods such as support vector machines (SVM), neural networks (NN), and deep learning (DL) architectures has made a significant contribution. Notably, these systems are extremely accurate at identifying fraud and forecasting stock values [14-15]. The landscape of stock prediction has transitioned from traditional statistical models like autoregressive integrated moving averages (ARIMA) and generalized autoregressive conditional heteroskedasticity (GARCH) towards machine learning and deep learning paradigms. These newer approaches exhibit superior capabilities in capturing intricate patterns within financial time series data. Widely utilized machine learning techniques for stock prediction encompass random forests,

SVMs, and gradient boosting machines (GBMs), effectively applied across various financial markets. Of particular note, deep learning models such as recurrent neural networks (RNNs) and long short-term memory (LSTM) networks have gained prominence for their adeptness in modelling temporal dependencies inherent in time series data [16-21]. Additionally, convolutional neural networks (CNNs), originally designed for image processing, have been repurposed for time series analysis and exhibit promising outcomes in the domain of stock prediction. The role of natural language processing (NLP) in these advancements, however, requires further elucidation for a clearer understanding.

*C. Blockchain technologies*

The rapid proliferation of blockchain technologies has prompted a transformative shift in the global financial landscape, with an increasing number of applications and platforms harnessing the benefits of decentralized and transparent ledgers to mitigate the risks associated with financial fraud. Several state-of-the-art AI-based solutions have been proposed to augment blockchain's fraud detection capabilities, with deep learning (DL) models, such as convolutional neural networks (CNNs) and recurrent neural networks (RNNs), demonstrating remarkable proficiency in identifying complex, multi-dimensional patterns within large-scale, heterogeneous dataset. Moreover, the application of unsupervised learning techniques, including clustering and dimensionality reduction, has engendered the identification of underlying structures and relationships among data points, thereby enabling the construction of more robust and generalizable models. Furthermore, the fusion of ML and natural language processing (NLP) technologies has facilitated the extraction of relevant information from unstructured data sources, such as social media platforms and news articles, thereby enriching the contextual understanding of transac-tions and their associated risk profiles.

*D. Sustainable finance and environmental, social, and governance (ESG)*

Sustainable finance has emerged as a critical topic in the contemporary financial ecosystem, especially in light of the increasing focus on environmental, social, and governance (ESG) factors. The burgeoning interest in sustainable finance has been paralleled by a surge in research related to financial fraud detection, which seeks to identify and prevent fraudulent activities that may undermine the integrity of financial markets and impede the realization of ESG goals. Financial fraud detection, a critical aspect of risk management, benefits from advancements in data analytics, machine learning, and artificial intelligence (AI) as they foster more accurate, efficient, and robust detection mechanisms. Recent literature highlights the growing importance of ESG disclosure and reporting, which enables investors and financial institutions to evaluate the non-financial performance of companies, identify potential risks, and assess their alignment with sustainability goals.

## V. Conclusion

The present study has analyzed the published literature on the topic of the application of AI & and ML in finance from 2010 to 2022 and critically demonstrated the findings in varied formats. The analysis of the publication data in this field over the years provides valuable insights into its growth and intellectual development. The increasing trend in the number of papers published from 2010 to 2022 signifies the growing interest and expanding research activity in the field, demonstrating its importance and relevance in the contemporary academic landscape. The oscillatory citation trajectory over the years highlights the varying influence of the research, suggesting potential resurgences and attenuations in interest, yet ultimately indicating the paper's academic significance. An examination of the research landscape reveals the noteworthy contributions of various authors and institutions in the field. The present research found that China is the most productive country in terms of several publications and the University of Oxford is the most productive institution. Amongst the countries, China, the UK, the USA, and India are on the top list. Since the present study only considers the 2010–2022 period, several exemplary literature and scientific worlds might have been excluded from the present analysis. Additionally, the current study only considers the economics and business development category for filtering out the paper search process, so it is obvious that many other relevant literatures that were published in other domains had not been included in the analysis. I will employ a combination of advanced analytical techniques such as co-citation analysis, bibliographic coupling, and network analysis to interpret and synthesize the amassed data. These issues are planned to be addressed in future attempts.